\documentclass[11pt]{article}

\usepackage[dvipsnames,usenames]{color}

\usepackage{colortbl}
\usepackage{graphicx}
\usepackage{amssymb}
\usepackage{here}
\usepackage{cite}
\newcommand{\AmS}{{\protect\the\textfont2
  A\kern-.1667em\lower.5ex\hbox{M}\kern-.125emS}}
\newcommand{\ba}{\begin{array}}
\newcommand{\ea}{\end{array}}

\def\beq{\begin{equation}}
\def\eeq{\end{equation}}

\def\bea{\begin{eqnarray}}
\def\eea{\end{eqnarray}}

\def\beq{\begin{equation}}
\def\eeq{\end{equation}}

\def\bea{\begin{eqnarray}}
\def\eea{\end{eqnarray}}

\begin{document}
\begin{titlepage}

\begin{flushright}
Report: IFIC/12-03, FTUV-12-0131\\
\end{flushright}

\begin{center}
\vspace{2.cm}
{\Large{\bf Prime numbers, quantum field theory \\
\vskip 0.2cm
and the Goldbach conjecture}}
\\
\vspace{1.2cm}

{\bf{Miguel-Angel Sanchis-Lozano\footnote{Email:Miguel.Angel.Sanchis@ific.uv.es}} \\
\vskip 0.1cm
\it Instituto de F\'{\i}sica
Corpuscular (IFIC) and Departamento de F\'{\i}sica Te\'orica \\
\it Centro Mixto Universitat de Val\`encia-CSIC \\
Dr. Moliner 50, 46100 Burjassot, Valencia (Spain)

\vskip 0.4cm

{\bf J. Fernando Barbero G.\footnote{fbarbero@iem.cfmac.csic.es}}\\
\vskip 0.1cm
\it Instituto de Estructura de la Materia, CSIC \\
\it Serrano 123, 28006 Madrid (Spain)

\vskip 0.4cm

{\bf Jos\'e Navarro-Salas\footnote{Email:jnavarro@ific.uv.es}}
\vskip 0.1cm
\it Instituto de F\'{\i}sica
Corpuscular (IFIC) and Departamento de F\'{\i}sica Te\'orica \\
\it Centro Mixto Universitat de Val\`encia-CSIC \\
Dr. Moliner 50, 46100 Burjassot, Valencia (Spain)}

\vspace{0.5cm}
\today{}

\begin{abstract}

Motivated by the Goldbach conjecture in Number
Theory  and the abelian bosonization mechanism on a cylindrical two-dimensional spacetime we study the reconstruction of a
 real scalar field
as a product of  two real fermion (so-called \textit{prime}) fields whose
Fourier expansion exclusively contains prime modes.
We undertake the canonical quantization of such prime fields
and construct the corresponding Fock space by
introducing creation operators
$b_{p}^{\dag}$ --labeled by prime numbers $p$-- acting on the vacuum.
The analysis of our model, based on the standard rules of
quantum field theory and the assumption of the Riemann hypothesis, allow us to prove that the theory is not renormalizable.
We also comment on the potential consequences of this result concerning the validity or breakdown
 of the Goldbach conjecture for large integer numbers.

\end{abstract}

\end{center}
\end{titlepage}

\newpage


\begin{flushright}
{\em Omnia disce, videbis postea nihil esse superfluum}\\
(Learn everything, later you will see that nothing is superfluous.)\\
Hugh de Saint Victor (c. 1078-1141), French or German philosopher
\end{flushright}

\vskip 1.4cm

\section{Introduction}

In this paper, we focus on the border between physics
and number theory. It can be regarded as a first step to
investigate, in particular, whether quantum field theory can shed any
light on some long-standing, yet unproved, mathematical problems in
number theory.
Indeed it seems reasonable that the ``translation'' of certain difficult mathematical questions into the physical language might provide
essential clues to better understand (or maybe help to solve) them.
For instance, it is well-known that the Hamiltonian describing quantum-mechanically a one-dimensional harmonic oscillator has an equally spaced spectrum (in the appropriate units and after a shift corresponding to the zero point energy) given by the set of positive integers. This remarkable property naturally leads to the following question: Is it possible to use spectral properties of this or other operators to study and analyze problems of number theory, e.g. the distribution of primes or the Goldbach conjecture?

At the beginning of the 20th century, P\`{o}lya and Hilbert independently suggested the possibility of proving the Riemann hypothesis on the distribution of non-trivial zeroes of the Riemann function $\zeta(s)$ {by finding a Hermitian operator with eigenvalues that could be put in correspondence with those of $\zeta(s)$ \cite{Berry:1999}. Moreover, a remarkable connection between number theory and nuclear physics emerged (see \cite{Firk} for a review)
as a consequence of a fortuitous meeting at Princeton in the early's 1970s between the mathematician Hugh Montgomery and the physicist Freeman Dyson. It was then noticed that Random Matrix Theory (introduced in the Physical Sciences by Eugene Wigner in the fifties \cite{Wigner1,Wigner2}) could model apparently unrelated problems of both fields. The level spacing of heavy nuclei and the non-trivial zeroes of the Riemann zeta function showed a striking statistical similarity.

This example is not the only one to suggest a link between physics and prime numbers. In 1990 Bernard Julia
put forward the idea of a fictitious, non-interacting, boson gas \cite{Julia:1990} whose constituents were called {\em primons}. The partition function of the primon gas was found to be the Riemann-zeta function $\zeta(s)$. The idea of the Riemann gas can also be extended to fermions leading to a relatively simple partition function of the form $\zeta(s)/\zeta(2s)$. Furthermore, it has been argued that more intriguing relations exist between the Riemann gas, quark confinement and even string theory \cite{Julia:1994}. This is a clear example of how studies initially started as a rather unorthodox approach to the Riemann hypothesis can lead,
in turn, to new and eventually fertile perspectives in physics and vice versa. A thorough summary of
several suggestive relationships between different branches of physics and the Riemann hypothesis can be found in \cite{Schumayer}.

In this paper we develop a (quite simple)
physical model involving a free fermionic field defined on a cylindrical space-time manifold, with prime excitation modes only.
No relation with p-adic theories in physics \cite{Brekke:1993gf,Volovich:1987nq} has been envisaged in this work. Rather, one of our goals is to introduce prime numbers within the standard language of Quantum Field Theory (QFT) and investigate its primary consequences. In particular, we will  cast Goldbach's conjecture into the standard language of Quantum Field Theory (QFT). Interestingly, as will be shown, the consistency analysis for the underlying QFT ultimately involves the Riemann hypothesis in the computation of the renormalized vacuum energy.}

\newpage
\subsection{Goldbach's conjecture}

In June 1742 the mathematician Christian Goldbach wrote a letter
to his friend, Leonhard Euler, where he made his famous
conjecture about prime numbers which can be split into the following two statements:
\begin{flushright}
{\small Every even integer greater than 2 can be expressed as the sum of two primes}
\end{flushright}
\begin{flushright}
{\small Every odd integer greater than 5 can be expressed as the sum of three
primes}
\end{flushright}
The first statement is known as the {\em binary} or {\em strong} Goldbach's conjecture (GC).
The second is known as the {\em ternary} or {\em weak} GC. Note that if the strong GC is true then the weak GC is also true. If not otherwise stated, we will denote the strong GC as the GC for short.\footnote{The number 1 was considered as prime in Goldbach's time, but presently it is usually not. This issue is, in any case, irrelevant for our study.}

The weak GC was proved in 1923 by Hardy and Littlewood, for sufficiently large odd numbers as an ``asymptotic" result, assuming the truth of the Riemann hypothesis. In 1937 Vinogradov again proved
the weak conjecture for a sufficiently large (but indeterminate) odd number using analytic methods.
In 1966 Chen Jing-Run proved that every sufficiently large even number can be expressed as the sum of a prime and a number with no more than two prime factors. At the present moment, the GC has been checked
up to integers $\sim 10^{18}$ \cite{Oliveira} by using computers.

\subsection{Polignac's conjecture}

In 1849 Alphonse de Polignac (1849) made the general conjecture that there are infinitely many cases of two consecutive prime numbers such that their difference is any even number $n_{even}$. For $n_{even}=2$ it reduces to the twin prime conjecture. For $n_{even}=4$ there should be infinitely many cousin primes... Based on heuristic arguments, Hardy and Littlewood found in 1922 a law to estimate the density of twin primes \cite{Hardy:1979}, also discussed by Clement in 1949 \cite{Clement}.

The following weakened form of Polignac's conjecture (PC) can be enunciated by relaxing the condition on the necessity that both primes are consecutive:

\begin{flushright}
{\small Any even number can be expressed as the difference of two primes in infinitely many ways.}
\end{flushright}
Polignac's weakened conjecture can be seen as complementary to Goldbach's conjecture. Both Polignac's and Goldbach's conjectures have never been proved.

\subsection{Riemann hypothesis}

At first glance the Riemann hypothesis (likewise the two preceding conjectures) does not seem very complicated to state, but in fact it lies deep in the roots of number theory, in connection with the mysteries of the distribution of the prime numbers:

\begin{flushright}
{\small All non-trivial zeros of the zeta function have real part equal to one-half.}
\end{flushright}

The Prime Number Theorem is equivalent to the assertion that no zeros of $\zeta(s)$ on the complex plane ($s=\sigma+it$) lie on the boundary of the critical strip ($0 \leq \sigma \leq 1$). Understanding how the non-trivial zeros are located relative to the critical line ($\sigma=1/2$), remains today one of the deepest unsolved questions in Mathematics. Moreover, note that many developments in Number Theory are contingent upon the
validity of the Riemann hypothesis (RH). In this paper we explore the connection between the RH and the GC and PC on the light of a quantum field theoretical model to be developed below.

\section{Scalar and fermion fields on a cylindrical space-time}

Our starting point is the Lagrangian density of a classical free real scalar field $\Phi(t,x)$ in a two-dimensional spacetime:
\begin{equation}\label{freelag}
{\cal L}\ =\ \frac{1}{2}\ \partial_{\mu}\Phi(t,x)\ \partial^{\mu}\Phi(t,x) \ ,
\end{equation}
where $x$ denotes the  space coordinate, leading to the well-known massless Klein-Gordon
equation:
\beq\label{KG}
\partial_{\mu}\partial^{\mu}\ \Phi(t,x)\  \equiv\
\biggl[\frac{\partial^2}{\partial t^2}-
\frac{\partial^2}{\partial x^2}\biggr]\
\Phi(t,x)=0
\eeq
The inverse metric tensor $g^{\mu\nu}$ has been defined such that $g^{00}=1$, $g^{11}=-1$,
all other components vanishing.
In addition we will set $\hbar=c=1$ throughout the paper.
Since we have assumed a zero mass the theory inherits the two-dimensional conformal symmetry \cite{CFT}, given by the spacetime transformations
 $ x^{\pm}\to y^{\pm}(x^{\pm})$ with $x^{\pm}\equiv (t{\pm} x)/\sqrt{2}$.

 As we ultimately want to deal with integer numbers we will let the spatial dimension $\lq\lq$curling up into a circle'', becoming cyclic thereby introducing a discrete spectrum in the theory. In such $1+1$ dimensional space, the scalar field $\Phi(t,x)$ has to be periodic, satisfying the boundary condition:
\beq\label{period}
\Phi(t,x)=\Phi(t,x+2\pi R)
\eeq
Let us now expand this field as
\beq\label{fourier}
\Phi(t,x)=\frac{1}{\sqrt{2\pi R}}
\sum_{n=-\infty}^{\infty}\varphi_n^{}(t)e^{inx/R}
\eeq
where $\varphi_n^{}(t)$ are complex Fourier coefficients in the above expansion, corresponding to a given radius $R$. The coefficient in front of the sum is introduced for normalization purposes. Using the  equation of motion  (\ref{KG}) one can fix immediately the (normalized) positive frequency solutions
\beq \varphi_n^{}(t)= \frac{1}{\sqrt{2\omega_n}}e^{-i\omega_nt} \ , \eeq
where $\omega_n=|n|/R$ (for simplicity we  neglect the zero-mode solutions). The modes labeled by positive values of $n$ represent complex waves  that move from left to right, while negative values represent left-moving  modes. Obviously,
owing to the linear character of the Klein-Gordon equation (\ref{KG}),
the $\Phi(x,t)$ field can be split into left and right moving (chiral) fields,
$\Phi_{-}(t,x)$ and $\Phi_{+}(t,x)$, respectively, both obeying
the periodic constraint (\ref{period}), as:
\beq\label{split+-}
\Phi(t,x)=\Phi_{-}(t,x)+\Phi_{+}(t,x)
\eeq
in such a way that the former (latter) only contains (positive frequency) Fourier components labeled
by positive ($n>0$) and negative ($n<0$) integer  numbers, respectively. At the quantum level the Fourier expansion is re-expressed in terms of creation and annihilation operators
\bea\label{positive}
\Phi_{-}(t-x) & = & \frac{1}{\sqrt{2\pi R}}\ \sum_{n>0}\ \biggl[\frac{a^-_n}{\sqrt{2\omega_n}}
e^{-i(\omega_n t - nx/R)}\ +\frac{a^{- \dagger}_n}{\sqrt{2\omega_n}}e^{i(\omega_n t - nx/R)}\ \biggr]\\
\Phi_{+}(t+x) & = & \frac{1}{\sqrt{2\pi R}}\ \sum_{n<0}\ \biggl[\frac{a^+_n}{\sqrt{2\omega_n}}
e^{-i(\omega_n t - nx/R)}\ +\frac{a^{+ \dagger}_n}{\sqrt{2\omega_n}}e^{i(\omega_n t - nx/R)}\ \biggr] \ ,
\label{negative}
\eea
where $[a^-_n, a^{-\dagger}_m]=\delta_{nm}$, $[a^-_n, a^{-}_m]= 0 =[a^{-\dagger}_n, a^{-\dagger}_m]$, and also $[a^+_n, a^{+\dagger}_m]=\delta_{nm}$, $[a^+_n, a^{+}_m]= 0 =[a^{+\dagger}_n, a^{+\dagger}_m]$. The vacuum state $|0\rangle$ is defined by the relations $a^{\mp}_n|0\rangle=0$.

We go further and generalize the above splitting in terms of even and odd integer numbers
\beq\label{split}
\Phi(t,x)=\Phi^{(even)}(t,x)+\Phi^{(odd)}(t,x)
\eeq
with
\bea\label{evenodd}
\Phi^{(even)}(t,x) & = & \frac{1}{\sqrt{2\pi R}}\
\sum_{n_{even}}\ \biggl[\frac{a_n}{\sqrt{2\omega_n}}
e^{-i(\omega_n t - nx/R)}\ +\frac{a^{\dagger}_n}{\sqrt{2\omega_n}}e^{i(\omega_n t - nx/R)}\ \biggr]\\
\Phi^{(odd)}(t,x) & = & \frac{1}{\sqrt{2\pi R}}\ \sum_{n_{odd}}\ \biggl[\frac{a_n}{\sqrt{2\omega_n}}
e^{-i(\omega_n t - nx/R)}\ +\frac{a^{\dagger}_n}{\sqrt{2\omega_n}}e^{i(\omega_n t - nx/R)}\ \biggr] \ ,
\label{odd}
\eea
with analogous commutation relations for the creation and annihilation operators. From the point of view of CFT the decomposition (\ref{split}) is  somewhat bizarre, since it
partially breaks conformal invariance. Only the Virasoro generators $L_n= 1/2 \sum_{m_{even/odd}}: a_ma_{n-m}:$, with $n$ being an even integer, do preserve the above decomposition. However, in our considerations conformal symmetry will play a rather heuristic role.

Now, with the goal of restating the GC and the PC in a field-theoretic language, we write the even (positive) numbers as either the sum or the difference of two (non necessarily consecutive) primes $p$ and $q$:
\bea\label{goldbach2}
n_{even} & = & p+q \\
n_{even} & = & |p-q|
\eea
From Eq.(\ref{evenodd}) and taking into account the elementary property of the exponential function,
one could make the tentative hypothesis that the chiral field $\Phi_-^{(even)}(t,x)$
\beq
\Phi_{-}^{(even)}(t,x) =  \frac{1}{\sqrt{2\pi R}}\ \sum_{n_{even}>0}\ \biggl[\frac{a^{-}_n}{\sqrt{2\omega_n}}
e^{-in(t - x)/R}\ +\frac{a^{-\dagger}_n}{\sqrt{2\omega_n}}e^{in(t - x)/R}\ \biggr]
\eeq
can be reconstructed as a  composite field according to
$\Phi_{-}^{(even)}(t-x) \propto \phi_-(t-x)\phi_-(t-x)$
where the chiral field
$\phi_-(t-x)$ is expressed by definition as a Fourier expansion
over prime modes exclusively.
This ansatz does not fit smoothly with the underlying conformal symmetry. Neither $\Phi_-$ and $\phi_-$ have well-defined properties under conformal transformations. However, this proposal can be improved
on the basis of the classical formulae on the abelian bosonization in two-dimensions \cite{coleman, mandelstam}. It is well-known that the light-cone derivatives of the scalar field $\partial_{\pm}\Phi$ are conformal fields which can be reconstructed in terms of fermion bilinears, as we review next (see also \cite{GO}).

\subsection{Bosonization in two dimensions}

To describe massless fermions in two dimensions it is useful to use a real representation of $\gamma$-matrices
\beq  \gamma^0=\left(\begin{array}{cc}  0  &   1  \\
1 & 0 \end{array}\right) \ \ \ \ \ \        \gamma^1=\left(\begin{array}{cc}  0  &   -1  \\
1 & 0 \end{array}\right)    \ \ \ \ \ \   \gamma^5=\gamma^0\gamma^1=\left(\begin{array}{cc}  1  &   0  \\
0 & -1 \end{array}\right)                                 \eeq
and write  a Dirac field using two real components
\beq  \Psi=\left(\begin{array}{c}  \Psi_-  \\
\Psi_+ \end{array}\right) \ . \eeq
The equation of motion $\gamma^{\mu}\partial_{\mu} \Psi =0$ implies  immediately that  $\Psi_-=\Psi_-(t-x)$ and $\Psi_+=\Psi_+(t+x)$. Moreover, the theory turns out to be invariant under conformal transformations $ x^{\pm}\to y^{\pm}(x^{\pm})$, where the chiral Fermi fields transform as follows \beq
\Psi_{\pm}(y^{\pm})=\left( \frac{dx^{\pm}}{dy^{\pm}}\right) ^{1/2}
\Psi_{\pm}(y^{\pm}(x^{\pm})) \ . \eeq
This means that the conformal algebra for the Dirac field has  central charge $c=1/2$. It is possible to reconstruct a scalar field $\Phi$, which  has  central charge $c=1$, from two real Dirac fields $\Psi^i$, with $i=1,2$. The currents $\Psi_{\pm}^1\Psi_{\pm}^2$ acquire a simple form in terms of derivatives of $\Phi$
 \beq :\Psi_{\pm}^1\Psi_{\pm}^2:={\mp}\frac{1}{\sqrt{2\pi}}\partial_{\pm}\Phi \ , \eeq
 where one introduces normal ordering to avoid divergences. We will also assume for simplicity
that the fermionic fields
satisfy the same periodic conditions \footnote{Fermionic fields can also satisfy
anti-periodic conditions yielding half-integer modes in the Fourier expansion.}
of (\ref{period}), i.e. $\Psi^j_{\pm} (t, x + 2\pi R)=\Psi^j_{\pm}(t, x)$.

Expanding  the fermion fields in modes (for simplicity we restrict to the right-moving sector)
\beq\label{positive2}
\Psi^j_{-}(t-x)  =  \frac{1}{\sqrt{2\pi R}}\ \sum_{n}b^j_n
e^{-in(t - x)/R}
\ , \eeq
where $\{b^i_n, b^{j }_m\}= \delta^{ij}\delta_{n, -m}$, one recovers the scalar annihilation and creation operators as bilinears involving all fermion operators $b^i_r$ (we use the usual definitions $b^{i \dagger}_r= b^{i }_{-r}$
and $a^{-\dagger}_n=a_{-n}^-$)
\beq\label{eq:a}
a^-_n= sgn(n)\sum_r i :b^1_rb^2_{n-r}: \ , \eeq
with the commutation relations $[a_{n}^-, a^{-\dagger}_{m}]= \delta_{n,m}$, with $n>0$, or, equivalently
\beq
 [a_{n}^-, a_{-m}^-]= \delta_{n,m}
\ . \eeq

\subsection{The even scalar field in terms of $\lq\lq$odd'' fermionic fields}

The trivial fact that any even positive number can be written as a sum of two odd numbers
suggests the possibility of extending the reduction of an even boson field in terms of two odd fermion fields.  In fact, one can easily obtain  a relation (and a similar one for the opposite chirality) equivalent to Eq.(\ref{eq:a})
this time for entire $n$ such that:

\beq a^{-}_{n}= sgn(n)\sum_{r_{odd}} i :b^1_rb^2_{n-r}: \eeq
where $b^j_{r} $ in the above expression are  annihilation and creation operators associated with fermion fields with Fourier expansion {\it over odd modes exclusively}
 \beq \Psi_-^{(odd)j}= \frac{1}{\sqrt{2\pi R}}\ \sum_{n_{odd}}b^j_n
e^{-in(t - x)/R}
\ . \eeq
The difference with respect to the standard fields is that the commutation relations for the bosonic creation/annihitation  operators are now of the form (with $n$ and $m$ even entires):
\beq
 [a^-_{n}, a^-_{-m}]= \frac{1}{2}\delta_{n,m}
\ . \eeq

 \subsection{The even scalar field in terms of $\lq\lq$prime'' fermionic fields}

The above mechanism and the GC (and also PC) suggest to continue the reduction process and consider fermion fields $\psi^j$ whose Fourier expansion {\it exclusively contains (odd) prime modes}\footnote{From our point of view it is natural to consider the number one as a prime number. We denote by $\mathbb{P'}$ the set of all (odd) prime numbers, including the number one.}
 \beq \label{psi}\psi^j= \frac{1}{\sqrt{2\pi R}}\ \sum_{\ {\pm}p\in\mathbb{P'}}b^j_p
e^{-ip(t - x)/R}
\ . \eeq
The (even) bosonic field can be reconstructed from the expression
\beq a^-_{n}= sgn(n)\sum_{\ {\pm}r\in\mathbb{P'}} i :b^1_rb^2_{n-r}: \ , \eeq
and it has now  the following non-canonical commutation relations\footnote{One can recover standard commutation relations with a redefinition of the operators $a^{-}_{n}$.}
\beq
 [a^-_{n}, a^-_{-m}]= \frac{r(n)}{n}\delta_{n,m}\ ,
\eeq
with $n$ and $m$ even entires; $r(n)$ stands for the number of
times the positive even integer $n$ can be written as the sum of two primes (the so-called Goldbach's partition).

Notice that the Goldbach conjeture has been invoked in order to reconstruct the
field $\partial_-\Phi^{even}$ as a product of two fields $\psi^j$ with only prime Fourier components.
This becomes explicit in the assumed non-vanishing function $r(n)$ in the above commutation relations.

\subsection{Even and odd states as $\lq\lq$prime'' excitations of the vacuum}

According to quantum field theory, $b^j_p$ and $b_p^{j \dagger}\equiv b^j_{-p}$ can be interpreted as annihilation and creation operators acting on the vacuum state denoted as $| 0_R \rangle$
such that
\beq\label{vacuum}
b_p^j | 0_R \rangle=0
\eeq
Once the creation and annihilation operators  have been defined, one can readily obtain two-prime states as
\beq\label{2p}
|p_1,p_2\rangle\  =\  N\
b_{-p_1}^{1}\ b_{-p_2}^{2 }\ |0_R\rangle
\eeq
where $n_{even}=p_1+p_2$; and $N$ stands for the normalization of states: $N=1$ if $p_1 \neq p_2$,
or $N=1/\sqrt{2}$ if $p_1=p_2$.

Similarly, a three-particle state results from
\beq\label{3p}
|p_1,p_2,p_3\rangle\ =\ N\
b_{-p_1}^{i}\ b_{-p_2}^{j}\
b_{-p_3}^{k}\ |0_R\rangle
\eeq
with $N=1$ for $p_1 \neq p_2 \neq p_3,\dots$
satisfying $n_{odd}=p_1+p_2+p_3$.

The validity of the GC implies that every even state, whose $\lq\lq$mass'' is given by
$M_n=n_{even}/R$, can be interpreted as a two-particle (two-prime) state $|p_1,p_2\rangle$,
where $n_{even}=p_1 + p_2$.
Correspondingly, every odd state, with mass $M_{n_{odd}}=n_{odd}/R $ can be interpreted as either
a one-prime state (for $n_{odd}$ a prime number itself) or a three-prime state $|p_1,p_2, p_3\rangle$, where
$n_{even}=p_1 + p_2 + p_3$ \footnote{Actually, primes greater than 5 can be expressed as the sum of three primes
according to the ternary GC.}. Leaving aside the fact that we are dealing with a noninteracting field theory, states
such as the two- and three-prime excitations of the vacuum bear some resemblance to hadrons, i.e. mesons and baryons, interpreted in non-relativistic models of particle physics as quark-antiquark and three-quark (or antiquark) states, respectively.

Naturally, four-prime states (or higher) can be formed by acting four times with the creation operation on the vacuum:
\beq\label{4p}
|p_1,p_2,p_3,p_4\rangle\ =\ N\
b_{-p_1}^{i}\ b_{-p_2}^{j}\
b_{-p_3}^{k}\ b_{-p_4}^{l}\ |0_R\rangle
\eeq
and so on for any even combination of creation operators.

Given an even integer $n_{even}$ in our physical model, the GC implies that all those Fock states such that
$n_{even}=p_1+p_2=p_1'+p_2'+p_3'+p_4'=\cdots$, are in fact degenerate. (Of course, similar extensions can be applied to odd combinations of creation operators.) Would the GC fail for a given $n_{even}$, the corresponding state could not be represented as a two-prime state anymore but likely as a {\em higher} Fock state. In Particle Physics jargon, one would refer to such states as $\lq\lq$exotic''.

The particle interpretation of the model presented above requires a consistent definition of the vacuum energy on a two-cylindrical spacetime. We examine in detail this issue in the following sections. Specifically,
we will study the regularization and renormalizability properties of both a bosonic field
and a fermionic field on the circle, the latter
with vanishing Fourier coefficients for modes with non-prime labels. We will pay special attention to the two point functions and the expectation value of the Hamiltonian operator (Casimir energy).

\subsection{Vacuum zero-point energy}

The expectation value of the the Hamiltonian $H_{\Phi}$ of the free scalar field $\Phi(x,t)$,
can be written  as
\beq\label{sumharm}
\langle 0_R|H_{\Phi}| 0_R\rangle\ =\ \frac{1}{2}\sum_{n}w_n,\ \ w_n=|n|/R
\eeq
where the sum runs over all integer numbers and the vacuum state corresponds to a given radius $R$. This quantity obviously diverges as $n \to \infty$.

On the other hand, in the absence of gravity the zero-point energy actually has no physical meaning in itself. This motivates the removal of the infinite vacuum energy of free Minkowski space by defining the Casimir energy $E_C$ as in \cite{DeWitt75, birrel-davies,  Bordag:2001qi}:
\beq\label{Casimir}
E_C=\langle 0_R|H_{\Phi}|0_R \rangle - \lim_{R\to \infty}\langle 0_R|H_{\Phi}|0_R \rangle
\eeq

Nowadays it has become standard practice the use of a regularization method to determine the Casimir energy based
on the Riemann $\zeta$ function \cite{Elizalde:2003jw,Elizalde:1994gf} defined as
\beq\label{zeta}
\zeta(s)\ =\ \sum_{n}n^{-s}
\eeq

Analytic continuation of $\zeta(s)$ from the region of the complex plane such that $\rm{Re}(s) >1$, to the whole complex plane including the real negative axis (and therefore $s=-1$) permits the regularized summation
of the divergent series (\ref{sumharm}). Thus one easily gets the textbook result for the vacuum energy density $\rho_0$:
\begin{equation}
\rho_0\ =\ \frac{E_C}{2\pi R}\ =\ -\frac{1}{24\pi R^2}
\end{equation}
Moreover, the vacuum energy energy for the chiral fields $\Phi_{\pm}$ is, obviously, half of the vacuum energy density obtained above for $\Phi$.
We also note that the Casimir energy density when the sum over the modes is restricted to even (odd) integers is $-1/6\pi R^2$ ($1/12\pi R^2$) \footnote{The Ramanujan summation of the infinite series of even and odd integers:
\[ 2^{-s}+4^{-s}+\cdots=2^{-s}\zeta(s),\ \ \ \ \
1^{-s}+3^{-s}+\cdots=(1-2^{-s})\zeta(s)
\]
leads to $-1/6$ and $1/12$ for $s=-1$, respectively \cite{Hardy}.}.

\subsubsection{Fermionic fields}

For the massless Dirac field $\Psi$ considered before the expectation value of the Hamiltonian
 $H_{\Psi}$ is minus the corresponding scalar field result
\beq\label{sumharmspsi}
\langle 0_R|H_{\Psi}| 0_R\rangle\ =\ -\frac{1}{2}\sum_{n}w_n,\ \ w_n=|n|/R
\eeq
Therefore, the regularized vacuum energy density turns out to be positive, with value $1/24\pi R^2$.

On the other hand, we are considering $\psi^j(t,x)$ introduced in Eq.(\ref{psi}) as the fundament fields in our model. Consequently, the expectation value of the hamiltonian $H_{\psi}$  reads now
\beq\label{sumprimes}
\langle 0_R|H_{\psi}| 0_R\rangle\ =\ -\frac{1}{2}\sum_{\  p \in\mathbb{P'}}w_p,\  w_p=p/R
\eeq
where the sum runs over all (odd) prime numbers. Similarly to (\ref{sumharm}) and (\ref{sumharmspsi}), this quantity also diverges as $p \to \infty$ and requires a regularization procedure.

In fact, the formal similarity of summations (\ref{sumharm}) and (\ref{sumprimes}) suggests trying
a similar regularization as before but this time using the the Riemann Prime zeta function, defined as\footnote{The fact that $\mathbb{P}\cup \{1\}=\mathbb{P'}\cup \{2\}$ implies that any successful regularization that can obtained with $\mathbb{P}$ can also apply to $\mathbb{P'}$, and viceversa.}

\beq\label{P}
P(s)=\sum_{p \in\mathbb{P}}p^{-s} \ .
\eeq
Unfortunately, $P(s)$ cannot be analytically continued to the negative real half-plane $(\rm{Re}(s)<0)$
because of a natural boundary along the line $\rm{Re}(s) = 0$ \cite{Froberg}. Therefore, the above method to obtain the Casimir energy, though successful for integer modes as shown before, cannot be applied to our model
with prime modes only. This failure should not be seen as exceptional as the Casimir effect regularization based on the use of the analytic continuation of the Riemann function is known to be not always applicable. This happens, for instance, in the case of a massive field on a two-dimensional cylinder \cite{Kay79}. In the next section we will approach this important question by a more systematic method based on the point-splitting regularization and renormalization tools of quantum field theory \cite{birrel-davies, Wald, Parker-Toms}.

 \section{Quantum prime field theory}

 \subsection{Scalar field}

It is instructive to consider the calculation of the vacuum energy of the scalar field $\Phi(t,x)$ by means of the so-called Abel summation of the divergent series (\ref{sumharm}). To motivate this procedure we start with the two-point function of the field, given by the sum in modes
\beq
\langle 0_R|\Phi (t, x) \Phi (t', x')|0_R \rangle = \frac{1}{2\pi R}\sum_{n=1}^{\infty} \frac{1}{2w_n} [e^{-in[(t-t' -i\epsilon)-(x-x')]/R}+  e^{-in[(t-t' -i\epsilon)+(x-x')]/R}] \ ,
\eeq
where, as usual, the infinitesimal $i\epsilon$ has been appropriately inserted to make the two-point function well-defined. As  is customary in quantizing a field in two-dimensions, we have discarded the zero mode $n=0$ in the above summation. At coincidence $(t', x') \to (t, x)$ we have
\beq
\langle 0_R|\Phi^{2}(t, x)|0_R \rangle = \frac{1}{2\pi}\sum_{n=1}^{\infty} \frac{1}{n} e^{-\epsilon n/R} \ ,
\eeq
which turns out to be the Abel summation of a divergent series \cite{Hardy}. From the sum
\beq \sum_{n=1}^{\infty} n^{-1}e^{-\epsilon n/R}= -\log (1-e^{-\epsilon/R}) \ , \eeq one can easily derive the expression for the two-point function \cite{Kay79}
\beq \label{tpf}
\langle 0_R|\Phi (t, x) \Phi (t', x')|0_R \rangle = -\frac{1}{4\pi}\log(1-e^{-i(\Delta u -i\epsilon)/R})(1-e^{-i(\Delta v -i\epsilon)/R}) \ ,
\eeq
where it is understood that the limit $\epsilon \to 0^+$ should be taken and  $\Delta u = u-u':= (t-x)-(t'-x')$ and $\Delta v = v-v':=(t+x)-(t'+x')$.

According to the above discussion, the energy density of the field, which classically is given by
\beq
T_{tt}= \frac{1}{2}\biggl[ \biggl(\frac{\partial \Phi}{\partial t}\biggr)^2 + \biggr(\frac{\partial \Phi}{\partial x}\biggr)^2 \biggr] \ ,
\eeq
can be formally evaluated, at the quantum level, as
\beq
\langle 0_R|T_{tt}|0_R \rangle = \lim_{\epsilon \to 0^+} (2\pi R)^{-1}\sum_{n=1}^\infty w_ne^{-\epsilon n/R}
\ .
\eeq
In the limit $\epsilon \to 0^+$, the quantity
\beq \label{sumdamping}(2\pi R)^{-1}\sum_{n=1}^\infty w_ne^{-\epsilon n/R}= \frac{1}{2\pi R^2(e^{\epsilon/2R} - e^{-\epsilon/2R})^2} \ .
\eeq
is divergent, reflecting the well-known divergent vacuum energy.  However, as is usual in quantum field theory one can invoke {\it renormalization} to properly find a finite quantity for the physical vacuum energy density. To renormalize $\langle 0_R|T_{tt}|0_R \rangle $ one can perform an inverse radius expansion in
\beq \label{Texpansion}
\langle 0_R|T^{(\epsilon)}_{tt}|0_R \rangle :=(2\pi R)^{-1}\sum_{n=1}^\infty w_ne^{-\epsilon n/R}= \frac{1}{2\pi \epsilon^2}-\frac{1}{24\pi R^2}+ O(R^{-4})
\ ,
\eeq
and renormalize by subtracting, from the exact expression (\ref{sumdamping}), the terms in the expansion (\ref{Texpansion}) that diverge when $\epsilon \to 0^+$
\beq \label{Tren}
\langle 0_R|T_{tt}|0_R \rangle_{ren} = \lim_{\epsilon \to 0^+}\biggl[\frac{1}{2\pi R^2}\frac{1}{(e^{\epsilon/2R} - e^{-\epsilon/2R})^2} - \frac{1}{2\pi \epsilon^2}\biggr] = -\frac{1}{24\pi R^2} \ .
\eeq
In the CFT  language, this result is somewhat related to the calculation of the central charge $c=1$ of the scalar field \cite{CFT}.
\subsection{Scalar field with even/odd integer numbers modes}

It is now illustrative to consider the restriction of the modes of the original field $\Phi$, parameterized by $n$, to the  set of  even (or odd) integer numbers. In this case we have
\beq
\langle 0_R|T^{(even)}_{tt}|0_R \rangle =  (2\pi R)^{-1}\sum_{n=even}^\infty w_ne^{-\epsilon n/R}= \frac{1}{4\pi \epsilon^2}-\frac{1}{6\pi R^2}+ O(R^{-4})
 \ ,
\eeq
and, after renormalization, we get
\beq
 \langle0_R|T^{(even)}_{tt}|0_R \rangle_{ren} = -\frac{1}{6\pi R^2} \ .
 \eeq
Similarly, for the quantum field defined on the odd integer numbers modes one gets
\beq
\langle 0_R|T^{(odd)}_{tt}|0_R \rangle_{ren} = \frac{1}{12\pi R^2} \ .
\eeq
We reproduce in this way  the results obtained in the previous section with the zeta function regularization.

\subsection{Fermi field with (odd) prime  modes}

 Let us now extend the above analysis to the fields $\psi^j$ introduced in section $2$. Since, by definition, the (chiral) field $\psi$ (for simplicity we omit the index $j$) has only (odd) prime number modes in its Fourier expansion, the corresponding two-point function is  then given by
 \beq \label{tpfprimes}
\langle 0_R|\psi (t-x) \psi (t'-x')|0_R \rangle = \frac{1}{2\pi R}\sum_{p\in\mathbb{P'}}  e^{-ip[(t-t' -i\epsilon)-(x-x')]/R} \ .
\eeq
At coincidence, one should deal with the following Abel sum
 \beq
\langle 0_R|\psi^{2 (\epsilon)}|0_R \rangle = \frac{1}{2\pi R}\sum_{p\in \mathbb{P'}}  e^{-\epsilon p/R} \ ,
\eeq
 which, as expected, is divergent when $\epsilon \to 0^+$.  Moreover, in the calculation of the vacuum energy density one faces directly the Abel sum of all (odd) prime numbers
 \beq
\langle 0_R|T^{prime \ (\epsilon)}_{tt}|0_R \rangle = -\frac{1}{4\pi R^2}\sum_{p\in \mathbb{P'}} p\ e^{-\epsilon p/R} \ .
\eeq
In sharp contrast with the field $\Phi$, the restriction to the prime numbers turns the above sums into rather complicated functions of $\epsilon/R$.

Our next and crucial step is to carry out a renormalization procedure to properly associate  finite quantities to the above divergent sums. As in the previous analysis of the field $\Phi$, this requires the introduction of suitable asymptotic expansions for large radius.
The leading terms in the asymptotic expansions of $\langle 0_R|T_{tt}^{(\epsilon)}|0_R \rangle$ is essentially fixed by the corresponding ones in Minkowski space: $\langle 0_R|T_{tt}^{(\epsilon)}|0_R \rangle \sim 1/2\pi \epsilon^2 +...$\ .

However, the definition of $\psi$
requires of a finite radius $R$ and thus it is not possible to define $\psi$ in Minkowski space. Therefore, the asymptotic expansions needed to identify the subtracting terms cannot be based on the results in Minkowski space and one has to construct them according to new criteria that we investigate in the following. Since our considerations should necessarily involve some conventional but sophisticated mathematics we will study the above Abel sums on the more conventional set  $\mathbb{P}$, instead of $\mathbb{P'}$.
Inasmuch as the fundamental aspects of the renormalization process only involves the behavior of the Abel sum as $p \to \infty$, the shift from $\mathbb{P'}$ to $\mathbb{P}$ is irrelevant (one should add the quantity $1/4\pi R^2$ to the  result obtained with $\mathbb{P}$).

Let us now focus on the renormalization of the expectation value of the energy density. To perform renormalization we should expand
$
\langle 0_R|T^{prime \ (\epsilon)}_{tt}|0_R \rangle$
for large radius.
The desired expansion can be computed either directly or by differentiating the expansion of $\sum_{p\in \mathbb{P}} \frac{1}{p} e^{-ap}$  ($a:=\epsilon/R$) with respect to the parameter $a$ twice and taking into account the additional factor $-1/4\pi R^2$. The result turns out to be the same.
After an involved calculation it is possible to show that the asymptotic expansion of $\sum_{p\in \mathbb{P}} \frac{1}{p} e^{-ap}$ when $a \to 0$ satisfies (without invoking the Riemann hypothesis)
\beq \label{fernandoexpansion1}
\sum_{p\in \mathbb{P}} \frac{e^{-ap}}{p}  \sim \log(-\log a)+B_1+ o(1)
\  , \eeq
where $B_1$ is Mertens' constant. This result is heuristically suggested by Mertens' theorem, but a rigorous proof involves non-trivial computations. For more details of this and other calculations we refer the reader to the Appendix at the end of the paper.

Note that the growth of the first (and only) divergent term on the r.h.s. of Eq.(\ref{fernandoexpansion1})
is very slow, due to the natural decrease of the density of prime numbers as $p \to \infty$. To go further in the analysis we are forced  to know the subleading terms in the expansion (\ref{fernandoexpansion1}). For instance, terms of order $O(1/\log a)$ in the expansion of $\sum_{p\in \mathbb{P}} \frac{1}{p} e^{-ap}$  are converted, after derivation, into divergent terms.   As explained in the Appendix one can  obtain the following expansion if the RH is assumed to be true

\bea \label{fernandoexpansion2}
\sum_{p\in \mathbb{P}} \frac{1}{p} e^{-ap} \sim \log(-\log a)+B_1-\sum_{k=1}^\infty\frac{(-1)^k\Gamma^{(k)}(1)}{k}\frac{1}{(-\log a)^k}
 \ ,
\eea
where $\Gamma^{(k)}(z)$ stands for the $k$-order derivative of the $\Gamma$ function.
For example, the first five terms in this expansion are
\bea
\sum_{p\in \mathbb{P}} \frac{1}{p} e^{-ap} &\sim&\log(-\log a)+B_1-\frac{\gamma}{(-\log a)}-\frac{\pi^2+6\gamma^2}{12}\frac{1}{(-\log a)^2} \nonumber \\ &-&\frac{4\zeta(3)+\gamma \pi^2+2\gamma^3}{6}\frac{1}{(-\log a)^3}+\cdots
\eea
The expansion for $\langle 0_R|T^{prime \ (\epsilon)}_{tt}|0_R \rangle$ can be expressed then as
\beq \label{fernandoexpansion3}
\langle 0_R|T^{prime \ (\epsilon)}_{tt}|0_R \rangle \sim  \frac{-1}{4\pi \epsilon^2} \sum_{k=0}^\infty \frac{(-1)^k \Gamma^{(k)}(2)}{(-\log (\epsilon/R))^{k+1}}                                                          \ . \label{asymtoticfernando2}\eeq
This series has an infinite number of divergent terms (as $\epsilon \to 0^+$) and their sum does not converge. Therefore, one cannot define at this point a proper subtracting term to cancel out the divergences and, hence, one could be tempted to conclude that renormalization is not possible.

However, a more careful inspection shows that this situation could be conceivably improved. To this end let us go back to the origin of (\ref{fernandoexpansion2}). As explained in the Appendix, by accepting the truth of the {\it Riemann hypothesis} it is possible to derive the following asymptotic expansion

\beq\label{fernandoexpansion4}
\sum_{p\in \mathbb{P}} \frac{1}{p} e^{-ap} = \frac{1}{2}\log(1+(-\log a)^2)+B_1-\mathrm{Im} \int_0^{\infty} e^{i t \log a} \left(\Gamma(-i t)-\frac{i e^{-t}}{t}\right)dt+ o(a^{1/2-\varepsilon})\ ,
\eeq
that, by combining a further  expansion of the first term with the third, leads to (\ref{fernandoexpansion2}). Similarly, assuming the Riemann hypothesis, the asymptotic expansion for $\langle 0_R|T^{prime \ (\epsilon)}_{tt}|0_R \rangle$ can be cast in the form

\beq \label{expansiontprime}
\langle 0_R|T^{prime \ (\epsilon)}_{tt}|0_R \rangle :=\frac{-1}{4\pi R^2}\sum_{p\in \mathbb{P}} p e^{-\varepsilon p/R} =\frac{-1}{4\pi R^2}(F(\epsilon/R)+  o((\epsilon/R)^{-3/2-\varepsilon}))\ ,
\eeq
where all divergences in (\ref{fernandoexpansion3}) are captured by the single function $F(a) = o(a^{-2})$\beq
F(a)=-\mathrm{Im}\frac{1}{a^2}\int_0^\infty e^{it\log a}\Gamma(2-it)\mathrm{d}t \ .
\eeq
It is important to notice, however, that the terms of the form $o(a^{-3/2 -\varepsilon})$ in (\ref{expansiontprime}) are potentially divergent. One would need to know the form of the non-leading contributions in (\ref{expansiontprime}) to see  whether or not the asymptotic expansion can be expressed as a finite sum of hierarchical divergent terms. In order to do this it would be necessary to go beyond the asymptotic analysis presented in the Appendix and study in detail the contributions to the asymptotic expansions coming from all the singularities in the integrands of (\ref{intrep}) and (\ref{intrep2}). The very complicated analytic structure of the relevant functions precludes us from making a definitive statement but, the presence of an infinite number of branch points originating in the non trivial zeros of the zeta function, strongly suggests that an infinite number of counterterms would be necessary to subtract all the infinities appearing in the expectation value of the stress-energy tensor. This
would mean that the model is non-renormalizable.

\section{Conclusions and final comments}

Motivated by the Goldbach and Polignac conjectures in Number Theory and the classical results on abelian bosonization in two-dimensions we
study in this paper the ``factorization'' of a  real scalar field (whose Fourier expansion is made out of even modes)  as a product of  two  (so-called prime) fermion fields whose Fourier expansion exclusively contains prime modes. We have subsequently investigated the quantum properties of this theory by studying the  vacuum energy density. As expected, the short-distance behavior of the prime field is highly non-trivial and quite different from a conventional field theory on the cylinder. Due to the restriction to the prime modes, the short-distance singularity is no longer purely of the form $1/\epsilon$, but rather of form $1/(\epsilon \log \epsilon)+ ...$ This feature makes the calculation of the renormalized expectation value of the energy density quite involved.  The computation  has turned out to be very elusive since an infinite number of divergences emerges naturally,
showing the non-renormalizability of the underlying QFT.

Let us stress that the Riemann hypothesis plays an important role in the calculation of  the vacuum energy density. This is, in our opinion, a remarkable outcome of this work. The possibility of relating consistency properties of a particular field theory (such as the one considered here) to the Riemann hypothesis would definitely be very intriguing.

The non-renormalizability of the theory  --in the standard sense--
would suggest that there is some scale at which its properties change dramatically. For instance, the two-particle sector of the Fock space reflects the GC at low energies. Accordingly, the set of all two-particle states span all energies of the form $n_{even}/R$. The non-renormalizability of the theory could signal a breakdown of the GC above some number density scale $N/R$, acting as a sort of ultraviolet cut-off. By breakdown of the GC we should understand in this context
that infinitely many even numbers cannot be written as the sum of two primes.

On the other hand, since the value of the radius $R$ can be freely chosen, the ultraviolet cut-off --and therefore $N$-- should be in principle also as large as desired. This view actually suggests that the GC might be satisfied for arbitrarily large even numbers. One can envisage this possibility on deeper physical grounds by extending our study beyond QFT, e.g. to string-like models, where dualities involving the size of $R$ (like $T$-duality, see for example \cite{Z}) could allow one to exploit this fascinating idea.

Let us finally stress that no actual claim is made in this paper regarding the proof or disproof of the Goldbach conjecture. Our main result is to point out that a simple theory, a free field on a cylindrical spacetime with modes restricted to prime integers, is full of divergences, making impossible to properly define the vacuum energy. We believe that  the truth of the Goldbach conjecture may be deeply related with this result, but in a way not unraveled yet.

\section{Appendix: Asymptotic expansions}

\noindent We give here some basic results about the asymptotic behavior  as $a\rightarrow 0^+$ of the functions
\begin{eqnarray*}
&&f(a)=\sum_{p\in \mathbb{P}}^\infty \frac{e^{-a p}}{p}\label{inverseprimes}\\
&&g(a)=\sum_{p\in \mathbb{P}}^\infty p e^{-a p}\label{primes}
\end{eqnarray*}
where the sums are extended to the set of prime numbers $\mathbb{P}$. As a way to highlight the role of the Riemann conjecture in the final form of the asymptotic expansions we will use complex variable methods.

\subsection{Asymptotic expansion for $f(a)$}

The functional series
$$
\sum_{p\in \mathbb{P}}^\infty \frac{e^{-a p}}{p}
$$
is convergent for $a\in \mathbb{C}$ with $\rm{Re}(a)>0$ and uniformly convergent in every compact subset of this set, hence it defines a holomorphic function in the complex half plane $\{a\in \mathbb{C}: \rm{Re}(a)>0\}$. The best way to study the asymptotic behavior of $f$ as $a\rightarrow 0^+$ is to use Mellin transform methods.
In the present case the integral defining the Mellin transform converges if $\rm{Re}(s)>0$ and is given by
\begin{equation}
M[f;s]=\sum_{p\in \mathbb{P}}\frac{1}{p}M[e^{-a p};s]=\Gamma(s)P(s+1)
\label{Mellintransformoff}
\end{equation}
where $P$ denotes the prime zeta function. This function is defined, in analogy with the standard zeta function, by the functional series
$$
P(s):=\sum_{p\in \mathbb{P}}\frac{1}{p^s}\,.
$$
for complex values of $s$ with $\rm{Re}(s)>1$. The Mellin inversion formula provides us with the following integral representation for $f$.
\begin{equation}
f(a)=\frac{1}{2\pi i}\int_{c-i\infty}^{c+i\infty}a^{-s}\Gamma(s)P(s+1)\mathrm{d}s\,\label{intrep}
\end{equation}
with $c>0$. This representation is the main tool to obtain the sought for asymptotic formulas for $f$ and $g$.

By using the expansion (see for example \cite{Froberg})
\begin{equation}
P(s)=\sum_{k=1}^{\infty}\frac{\mu(k)}{k}\log \zeta(k s)\,,\label{Mobinv}
\end{equation}
that can be obtained by using M\"{o}bius inversion\footnote{Here $\mu(k)$ denotes the M\"{o}bius function and $\zeta$ is the standard zeta function.}, we get
$$
f(a)=\frac{1}{2\pi i}\int_{c-i\infty}^{c+i\infty}a^{-s}\Gamma(s) \sum_{k=1}^{\infty}\frac{\mu(k)}{k}\log \zeta(k (s+1))\,,
$$
that can be equivalently written as

\begin{eqnarray}
f(a)&=&\frac{1}{2\pi i}\int_{c-i\infty}^{c+i\infty} a^{-s} \Gamma(s)\log(s\zeta(s+1))\mathrm{d}s-\frac{1}{2\pi i}\int_{c-i\infty}^{c+i\infty} a^{-s} \Gamma(s)\log(s)\mathrm{d}s\nonumber\\
&+&\frac{1}{2\pi i}\sum_{k=2}^\infty \frac{\mu(k)}{k}\int_{c-i\infty}^{c+i\infty}a^{-s}\Gamma(s)\log\zeta(k(s+1))\mathrm{d}s\,.
\label{tresintegrales}
\end{eqnarray}
for an appropriately chosen value of $c$.

\bigskip

Let us start by considering the first integral in (\ref{tresintegrales}). Without using the Riemann hypothesis this integral can be easily seen to be $O(1)$ by moving the integration contour to the line $\rm{Re}(s)=0$. This is possible because there are no zeros of the $\zeta$ function on the line $\rm{Re}(s)=1$. Also, we have $\Gamma(c+it)=O(t^{c-1/2}e^{-\pi t/2})$ and $\log|s\zeta(s+1)|=O(\log|t|)$ and, hence, the contributions of the segments needed to close the integration contour and displace it to the line $\rm{Re}(s)=0$ go to zero as $L\rightarrow\infty$. The previous result can be improved by noting that the Riemann-Lebesgue lemma actually shows that the limit of the integral when $a\rightarrow 0$ vanishes [or, equivalently, the integral is $o(1)$]. If the Riemann hypothesis is true the integral can be shown to be $O(a^{1/2-\epsilon})$ for all $\epsilon>0$ by moving the contour to the line $\rm{Re}(s)=-1/2+\epsilon$. Again this result can be slightly improved and the
 integral shown to be $o(a^{1/2-\epsilon})$ as a consequence of the Riemann-Lebesgue lemma.

\bigskip

Let us consider now the integrals appearing in
$$
\frac{1}{2\pi i}\sum_{k=2}^\infty \frac{\mu(k)}{k}\int_{c-i\infty}^{c+i\infty}a^{-s}\Gamma(s)\log\zeta(k(s+1))\mathrm{d}s\,.
$$
In each of the integrands, the singularities with real parts larger\footnote{The only ones relevant for the present analysis.} than -1 have several different origins:

\begin{itemize}
\item The ones with the largest real part correspond to the pole of $\Gamma$ at $s=0$. The residues are $\log \zeta(k)$ for $k=2,3,\ldots$.
\item The pole of $\zeta$ gives rise to branch points at $s=-1+1/k$, with $k=2,3,\ldots$.

\item The non-trivial zeros of the $\zeta$ function, if the Riemann hypothesis is true, give rise to branch points with
$\rm{Re}(s)=-1+\frac{1}{2k}$, and $k=2,3,\ldots$
\end{itemize}
As in the integral considered in the first step, the integration contour can be moved parallel to itself due to the exponential fall-off of the integrand and the mild growth of the absolute value of the $\zeta$ function. By moving the contour past the origin we get contributions, from the residues at the poles of the $\Gamma(s)$ factor, that add up to the constant $$C:=\sum_{k=2}^\infty \frac{\mu(k)}{k} \log \zeta(k)\,.$$
This, combined with the Euler constant $\gamma$, will give a final contribution equal to the Mertens constant $B_1=0.26149721\cdots$. Once past the $s=0$ singularity, the integration contour can be displaced to the line $c=-1/2+\varepsilon$ and it is straightforward to show that the sum of the integrals on this contour is $o(a^{1/2-\epsilon})$ without invoking the Riemann hypothesis.

\bigskip

Finally let us consider the integral
\begin{equation}
-\frac{1}{2\pi i}\int_{c-i\infty}^{c+i\infty} a^{-s} \Gamma(s)\log s\,\mathrm{d}s\,.
\label{f1}
\end{equation}

\begin{figure}[htbp]
\hspace{3cm}\includegraphics[width=10cm]{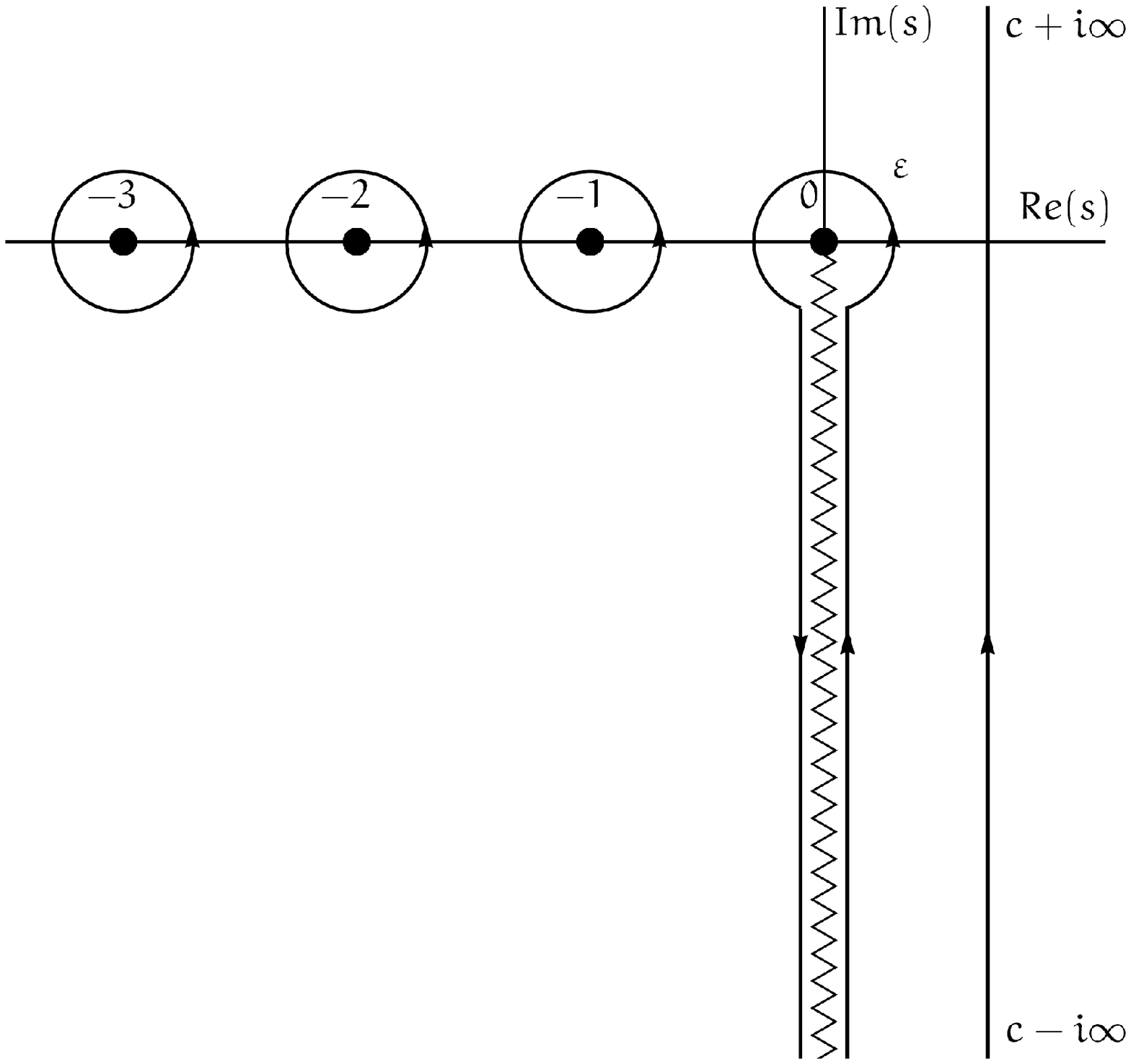}
\caption{Integration contour for the integral (\ref{f1}) used in the study of $f(a)$.}
\label{Fig:contour}
\end{figure}

The singularities in the integrand are determined now by the poles of the gamma function at the non-positive integers, and the branching point of the logarithm at $s=0$. The quick fall-off of the integrand provided by the gamma function allows us to freely move the integration contour in the negative direction of the real axis while wrapping around the singularities of the integrand. In order to compute the integral it is useful to somehow disentangle the singularity structure by choosing a branch cut for the logarithm such that $\arg(z)\in(-\pi/2,3\pi/2]$ (see figure \ref{Fig:contour}). As long as the choice does not change the value of the integrand in the region where the initial integration contour lies  this is allowed. The advantage of proceeding in this way is that the poles of the gamma function will not lie on top of the branch cut.

A straightforward computation gives now
\begin{eqnarray}
\mathrm{Re}\left(-\frac{1}{2\pi i}\int_{c-i\infty}^{c+i\infty}a^{-s}\Gamma(s)\log s \mathrm{d}s\right)&=&\gamma+\frac{1}{2}\log(1+(-\log a)^2)\label{re1}\\
&&-\mathrm{Im}\int_0^\infty\!\!\!\! e^{it\log a}\big(\Gamma(-it)-\frac{ie^{-t}}{t}\big)\mathrm{d}t+O(a^2)\,.\nonumber
\end{eqnarray}
It is important to realize that despite the relative complication of (\ref{re1}) it is actually easy to find its asymptotic behavior when $a\rightarrow 0^+$. In particular, the last integral can be easily seen to be $O(1/(-\log a))$ by integration by parts and its full asymptotic expansion can be obtained directly by this procedure. This gives
\begin{eqnarray*}
&&-\mathrm{Im}\int_0^\infty e^{i t\log a}\big(\Gamma(-i t)-\frac{ie^{-t}}{t}\big)\mathrm{d}t\sim -\mathrm{Im}\sum_{k=0}^\infty
\frac{k!e^{-\pi i (k+1)/2}}{(-\log a)^{k+1}}[t^k]\big(\Gamma(-i t)-\frac{ie^{-t}}{t}\big)\\
&&=-\sum_{k=1}^\infty\frac{(-1)^k\Gamma^{(k)}(1)}{k}\frac{1}{(-\log a)^{k}}+\frac{1}{2}\sum_{k=1}^\infty\frac{(-1)^k}{k}\frac{1}{(-\log a)^{2k}}\,.
\end{eqnarray*}
In the limit $a\rightarrow 0^+$ the asymptotic behavior is controlled by the divergence of $\frac{1}{2}\log(1+(-\log a)^2)$ that behaves as $\log(-\log a)$ for small enough values of $a$.
$$
\frac{1}{2}\log(1+(-\log a)^2)=\log(-\log a)-\frac{1}{2}\sum_{k=1}^\infty \frac{(-1)^k}{k}\frac{1}{(-\log a)^{2k}}\,.
$$

By taking together the results obtained for each of the integrals in (\ref{tresintegrales}) we conclude that, without invoking the Riemann hypothesis, we have
\begin{equation}
f(a)=\log(-\log a)+B_1+o(1)\,,
\label{as-exp}
\end{equation}
whereas, if the Riemann hypothesis is true we have
$$
f(a)=\frac{1}{2}\log\big(1+(-\log a)^2\big)+B_1-\mathrm{Im}\int_0^\infty\!\!\!\! e^{it\log a}\big(\Gamma(-it)-\frac{ie^{-t}}{t}\big)\mathrm{d}t+o(a^{\frac{1}{2}-\varepsilon})
$$
for every $\varepsilon>0$. Here $B_1:=\gamma+C$ is the Mertens constant mentioned above. As a side remark we would like to mention here that the asymptotic expansion (\ref{as-exp}) is the same \cite{Hardy:1979} as the one corresponding to the truncated series
$$
\sum_{p\in\mathbb{P};\,p\leq\frac{1}{a}}\frac{1}{p}=\log(-\log a)+B_1+o(1)
$$
in the limit $a\rightarrow 0$.

\subsection{Asymptotic expansion for $g(a)$}

The function $g(a)$ can be studied by following the same steps as before. Now the functional series
$$
\sum_{p\in \mathbb{P}}^\infty p e^{-a p}$$
is convergent for $a\in \mathbb{C}$ with $\rm{Re}(a)>0$ and uniformly convergent in every compact subset of this set, hence it defines a holomorphic function in the complex half plane $\{a\in \mathbb{C}: \rm{Re}(a)>0\}$. It is straightforward to find the following representation for $g(a)$ by using Mellin transforms.
\begin{equation}
g(a)=\frac{1}{2\pi i}\int_{c-i\infty}^{c+i\infty}a^{-s}\Gamma(s)P(s-1)\mathrm{d}s\,\label{intrep2}
\end{equation}
with $c>2$. This expression can be written as
$$
g(a)=\frac{1}{2\pi i}\int_{c-i\infty}^{c+i\infty}a^{-s}\Gamma(s) \sum_{k=1}^{\infty}\frac{\mu(k)}{k}\log \zeta(k (s-1))\,.
$$
The singularity structure of the integrand suggests to split the previous expression for $g(a)$ as
\begin{eqnarray}
g(a)&=&\frac{1}{2\pi i}\int_{c-i\infty}^{c+i\infty} a^{-s} \Gamma(s)\log((s-2)\zeta(s-1))\mathrm{d}s-\frac{1}{2\pi i}\int_{c-i\infty}^{c+i\infty} a^{-s} \Gamma(s)\log(s-2)\mathrm{d}s\nonumber\\
&+&\frac{1}{2\pi i}\sum_{k=2}^\infty \frac{\mu(k)}{k}\int_{c-i\infty}^{c+i\infty}a^{-s}\Gamma(s)\log\zeta(k(s-1))\mathrm{d}s\,.\label{intsforg}
\end{eqnarray}
for a certain value of $c$.

\bigskip

Without invoking the Riemann hypothesis the first integral in (\ref{intsforg}) can be easily shown to be $O(a^{-2})$ by moving the integration contour; actually, by invoking the Riemann-Lebesgue lemma it can be seen to be $o(a^{-2})$. If the Riemann hypothesis is true then this integral can be seen to be $O(a^{-3/2-\varepsilon})$ (a bound that can, again, be slightly improved to $o(a^{-3/2-\varepsilon})$ by using the Riemann-Lebesgue lemma).

Let us consider now the sum of integrals
$$
\frac{1}{2\pi i}\sum_{k=2}^\infty \frac{\mu(k)}{k}\int_{c-i\infty}^{c+i\infty}a^{-s}\Gamma(s)\log\zeta(k(s-1))\mathrm{d}s\,.
$$
At variance with the analysis presented for $f(a)$, in this case the singularities of $\Gamma(s)$ do not play any role to determine the asymptotic behavior of $g(a)$.  This expression can be seen to be $o(a^{-3/2-\varepsilon})$ regardless of the truth of the Riemann hypothesis.

\begin{figure}[htbp]
\hspace{3cm}\includegraphics[width=10cm]{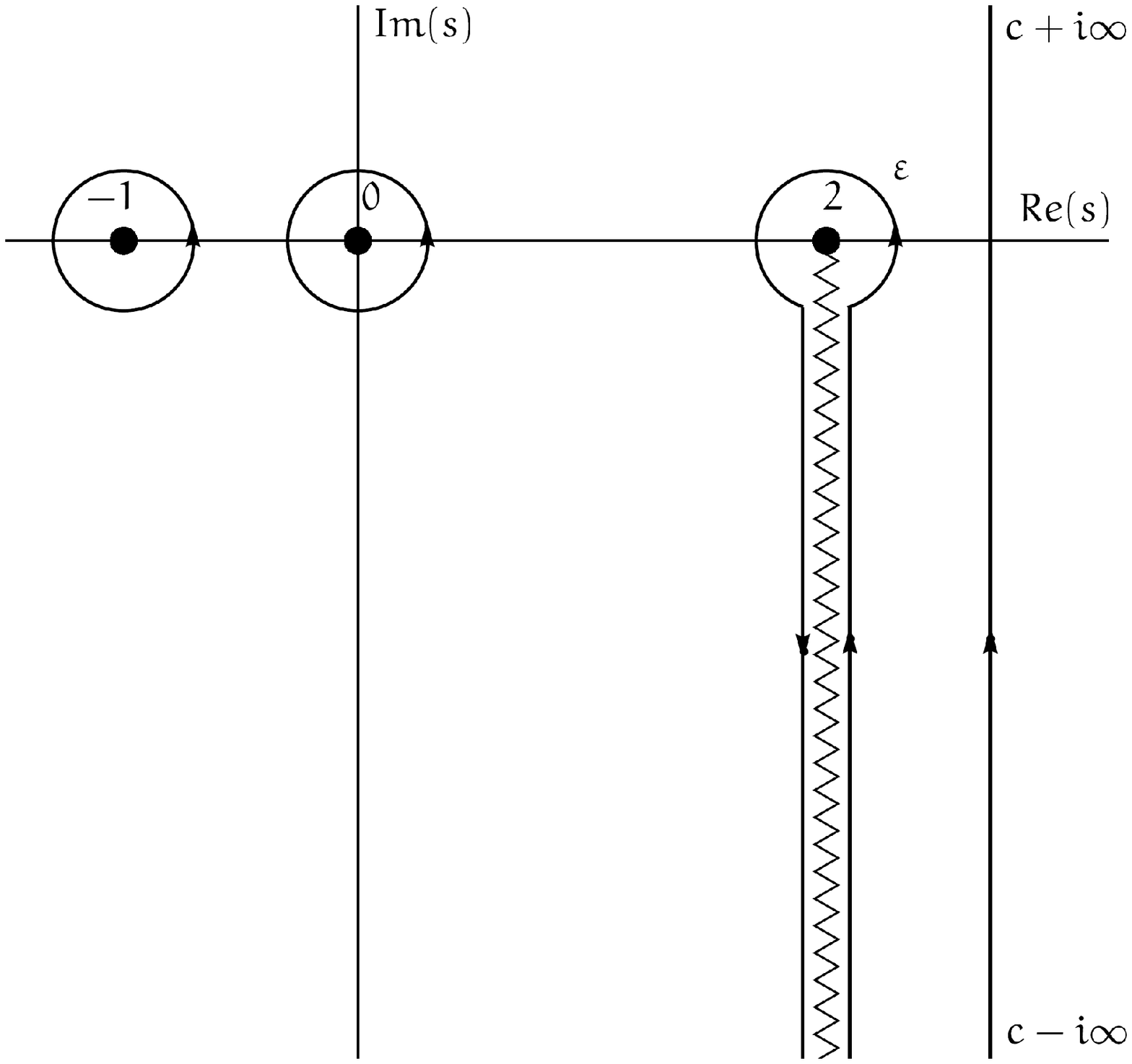}
\caption{Integration contour for the integral (\ref{f2}) used in the study of $g(a)$.} \label{Fig:contour2}
\end{figure}

Finally let us study the behavior of
\begin{equation}
-\frac{1}{2\pi i}\int_{c-i\infty}^{c+i\infty} a^{-s} \Gamma(s)\log(s-2)\mathrm{d}s\,.
\label{f2}
\end{equation}
The singularities of the integrand and the the integration contours used to study the asymptotic behavior of this integral are shown in Fig. \ref{Fig:contour2}. A simple computation gives now
\begin{eqnarray*}
\mathrm{Re}\left(-\frac{1}{2\pi i}\int_{c-i\infty}^{c+i\infty}a^{-s}\Gamma(s)\log (s-2) \mathrm{d}s\right)&=&-\frac{1}{a^2}\mathrm{Im}\int_0^\infty\!\!\!\! e^{it\log a}\Gamma(2-it)\mathrm{d}t+O(1)\,.\hspace*{18mm}
\end{eqnarray*}
The asymptotic behavior of this last integral can be easily obtained by integration by parts
$$
-\frac{1}{a^2}\mathrm{Im}\int_0^\infty\!\!\!\! e^{it\log a}\Gamma(2-it)\mathrm{d}t\sim -\frac{1}{a^2}\mathrm{Im}\sum_{k=0}^\infty\frac{k!e^{-\frac{\pi i (k+1)}{2}}}{(-\log a)^{k+1}}[t^k]\Gamma(2-i t)=\frac{1}{a^2}\sum_{k=0}^\infty \frac{(-1)^k\Gamma^{(k)}(2)}{(-\log a)^{k+1}}\,.
$$

Gathering the preceding results we conclude that without invoking the Riemann hypothesis we can only state that
$$
g(a)=o(a^{-2})\,,
$$
whereas, if the Riemann hypothesis is true, we have that
$$
g(a)=-\mathrm{Im}\frac{1}{a^2}\int_0^\infty e^{it\log a}\Gamma(2-it)\mathrm{d}t+o(a^{-\frac{3}{2}-\varepsilon})
$$
for all $\varepsilon>0$.

It is interesting to point out here that the leading asymptotics of $g(a)$ can be obtained by differentiating twice the asymptotic behavior of the leading terms in $f(a)$, i.e.
$$
\frac{1}{2}\log\big(1+(-\log a)^2\big)+B_1-\mathrm{Im}\int_0^\infty\!\!\!\! e^{it\log a}\big(\Gamma(-it)-\frac{ie^{-t}}{t}\big)\mathrm{d}t\sim
$$
$$
\log(-\log a)+B_1-\sum_{k=1}^\infty\frac{(-1)^k\Gamma^{(k)}(1)}{k}\frac{1}{(-\log a)^k}\,.\hspace*{2cm}
$$

\subsection*{Acknowledgments}
This work is supported by the MEYC research grants FPA2011-23596, FIS2009-11893, FIS2008-06078-C03-02,
the  Consolider-Ingenio 2010 Program CPAN (CSD2007-00042) and the Generalitat Valenciana grant GVPROMETEO2010-056. We thank L.~\'Alvarez-Gaum\'e and E.~Elizalde for useful comments.


\end{document}